\def\Roman#1{\uppercase\expandafter{\romannumeral#1}}
\documentclass[a4paper,12pt]{article}
\usepackage{amssymb,amsfonts}
\usepackage[mathscr]{eucal}

\title{Wong's equations in Yang-Mills theory}

\author{S.N.Storchak}

\author{S. N. Storchak\footnote{E-mail adress: storchak@ihep.ru}\\
\small{Institute for High Energy Physics, Protvino, Moscow Region,142284,Russia}}

\begin{document}

\maketitle

\begin{abstract}
We derive Wong's equations for the finite-dimensional  dynamical system representing the motion of a scalar particle  on a compact Riemannian manifold with a given free isometric smooth action of a compact semisimple Lie group. The obtained equations are  written  in terms of  dependent coordinates which are typically used in an  
implicit description of the local dynamics given on the orbit space of the principal fiber bundle. 
Using these equations we obtain Wong's equations  in  a pure Yang--Mills gauge theory with the Coulomb gauge fixing.
This result is based on the existing analogy between the reduction procedures carried out in our finite-dimensional dynamical system and in  Yang-Mills gauge fields.
\end{abstract}

\section{Introduction}
In his paper \cite{Wong}, S.K.Wong suggested   the equations for the phenomenological description of the strong interactions. His idea was to present this interaction by means of the classical motion of some particles. 
A new approach has given rise to an alternative method of studying the behavior at short distances in QCD theory.
(As an example of such an approach we refer to
\cite{x,y}.) 

We note that  one of Wong's  equations, the equation for the motion of the particle  in an external field, was earlier obtained by Kerner in \cite{Kerner}. 

Later Montgomery discovered \cite{Montgomery},  that Wong's equations are related to the dynamical systems with a symmetry. One can meets with the equations when 
considering  the reduction problems in  these systems. Wong's equations are the 
 horizontal and vertical parts of the geodesic equation written for the Kaluza-Klein metric
\cite{Marsd}.

In the cotangent bundle reduction problem,  Wong's equations
can be obtained by extracting the horizontal part of $ker(d\omega)$, where $\omega$ is  a connection one-form  which characterise the dynamical system \cite{Duval}. 

Although Wong's equations, 
resulting in reduction problems, coinside in form with the original Wong's equations (and with the equations that are used in applications), they have a different meaning. In the original equations, the field strength is a curvature of an arbitrary given gauge potential, whereas in the reduction problems, the field strength is a 
curvature of a special connection one-form, called
the mechanical connection.\footnote{This connection is naturally arises in the reduction procedure.} That is, the field strength is related to an effective internal interaction which exists in each dynamical system with a symmetry. 

 In  Yang--Mills theory, we are dealing with  
an infinite-dimensional dynamical system
 with  gauge degrees of freedom. 
In this theory,  a description of a local dynamics   can only be done in an implicit way. ``True motion'' of the system on the gauge orbit space  can be studied  
by considering the  evolution of the corresponding dynamical system given on the gauge surface. 
Since this surface is determined by the gauge conditions, this means that one should use the dependent variables to describe the evolution.

An appropriate method of such a description of the reduced motion  
in the finite-dimensional dynamical systems was previously developed  in our papers \cite{Storchak_1, Storchak_2}. 
 We considered the motion of the scalar particle on the compact Riemannian manifold on which an isometric smooth action of the semisimple compact Lie group was given. It was assumed that this action is free and proper. 
 As in gauge theories,  the reduced evolution in this system is defined   on the orbit space of a  group action.
 
The first part of our paper will be devoted to derivation of Wong's equations for our finite-dimensional dynamical system.  

In the second part of the paper,  after short reviewing of the reduction problem in a pure Yang-Mills theory, in which the reduced surface is given by means of the Coulomb gauge condition, we obtain Wong's equations for  Yang-Mills gauge theory.
This is done by generalizing Wong's equations for the finite-dimensional system.

In the last section we discuss some questions that must be addressed in future studies.

 Details of the derivation of Wongs's equations  are considered in Appendix.

\section{Definitions}
In this section we introduce the notations from our  papers 
 \cite{Storchak_1, Storchak_2}, where
the motion of the scalar particle on a compact manifold $\cal P$ was studied. 
A free  isometric smooth action\footnote{We consider the right action of the group on $\cal P$.} of a semisimple compact Lie group $\cal G$ on this manifold 
leads to   the fiber bundle picture with $\cal P$ as the total space
of this principal fiber bundle $\pi : \cal P \to {\cal P}/{\cal G}=\cal M$.

To perform the reduction procedure we must first replace the original coordinates $Q^A$,  given on a local chart of the manifold $\cal P$, by new coordinates 
$(Q^{\ast}{}^A,a^{\alpha})$ ($A=1,\ldots , N_{\cal P},N_{\cal P}=\dim {\cal P} ;{\alpha}=1,\ldots , N_{\cal G},N_{\cal G}=\dim {\cal G}$)
  related to the fiber  bundle. 
We are forced to introduce additional constraints ${\chi}^{\alpha}(Q^{\ast})=0$ in order to have  one-to-one correspondence between  the old coordinates $Q^A$ and $(Q^{\ast}{}^A,a^{\alpha})$.

These constraints define the local submanifolds in the  manifold $\cal P$.
If these local submanifolds (local sections) can be `glued' into the global manifold $\Sigma $, we get
   a trivial principal fiber bundle $P({\cal M},\cal G)$ which
is locally isomorphic to the trivial bundle 
$\Sigma\times {\cal G}\to{\Sigma} $. Therefore, we can  use the coordinates $Q^{\ast}{}^A$ for the description of the evolution which is given on the manifold $\cal M$.

Replacing the original coordinate basis $(\frac{\partial}{\partial Q^A})$ for 
a new coordinate basis $(\frac{\partial}{\partial Q^{\ast}{}^A},\frac{\partial}{\partial a^{\alpha}})$, we come to the following representation for  the original metric ${\tilde G}_{\cal A\cal B}(Q^{\ast},a)$ of the manifold $\cal P$: 
\begin{equation}
\left(
\begin{array}{cc}
G_{CD}(Q^{\ast})(P_{\perp})^{C}_{A}
(P_{\perp})^{D}_{B} & G_{CD}(Q^{\ast})(P_{\perp})^
{D}_{A}K^{C}_{\mu}\bar{u}^{\mu}_{\alpha}(a) \\
G_{CD}(Q^{\ast})(P_{\perp})^
{C}_{A}K^{D}_{\nu}\bar{u}^{\nu}_{\beta}(a) & {\gamma }_{\mu \nu }
(Q^{\ast})\bar{u}_\alpha ^\mu (a)\bar{u}_\beta ^\nu (a)
\end{array}
\right),
\label{1}
\end{equation}
 where $K_{\mu}$ are the Killing vector fields for the Riemannian metric 
$G_{AB}(Q)$.
(This vector fields are restricted to the submanifold $\Sigma \equiv\{{\chi}^{\alpha}=0\}$.)

In (\ref{1}), 
by ${\gamma}_{\mu \nu}$ we denote 
the metric given on the orbit of the group action. It is  defined by the relation ${\gamma}_{\mu \nu}
=K^{A}_{\mu}G_{AB}K^{B}_{\nu}$.

${\bar u}^{\alpha}_{\beta}(a)$ (and ${u}^{\alpha}_{\beta}(a)$) are the coordinate representations of the auxiliary functions given on the group $\cal G$.

The operator $P_{\perp}(Q^{\ast})$ is used to  projects the vectors onto the tangent space to the gauge surface $\Sigma$:
\[
(P_{\perp})^{A}_{B}=\delta ^{A}_{B}-{\chi}^{\alpha}_{B}
(\chi \chi ^{\top})^{-1}{}^{\beta}_{\alpha}(\chi ^
{\top})^{A}_{\beta},
\]
 $(\chi ^{\top})^{A}_{\beta}$ is a transposed matrix to the matrix $\chi ^{\nu}_{B}\equiv \frac{\partial \chi ^{\nu}}{\partial Q^B}$, 
$(\chi ^{\top})^{A}_{\mu}=G^{AB}{\gamma}_
{\mu \nu}\chi ^{\nu}_{B}.$

The pseudoinverse matrix ${\tilde G}^{\cal A\cal B}(Q^{\ast},a)$ to the matrix (\ref{1}), i.e.  such a matrix for which 
\begin{eqnarray*}
\displaystyle
{\tilde G}^{\cal A\cal B}{\tilde G}_{\cal B\cal C}=\left(
\begin{array}{cc}
(P_{\perp})^A_C & 0\\
0 & {\delta}^{\alpha}_{\beta}
\end{array}
\right),
\end{eqnarray*}  
is given by 
\begin{equation}
\displaystyle
\left(
\begin{array}{cc}
G^{EF}N^{C}_{E}
N^{D}_{F} & G^{SD}N^C_S{\chi}^{\mu}_D
(\Phi ^{-1})^{\nu}_{\mu}{\bar v}^{\sigma}_{\nu} \\
G^{CB}{\chi}^{\gamma}_C (\Phi ^{-1})^{\beta}_{\gamma}N^D_B
{\bar v}^{\alpha}_{\beta} & G^{CB}
{\chi}^{\gamma}_C (\Phi ^{-1})^{\beta}_{\gamma}
{\chi}^{\mu}_B (\Phi ^{-1})^{\nu}_{\mu}
{\bar v}^{\alpha}_{\beta}{\bar v}^{\sigma}_{\nu}
\end{array}
\right).
\label{2}
\end{equation}
The matrix  $(\Phi ^{-1}){}^{\beta}_{\mu}$  is inverse
to the Faddeev -- Popov matrix $\Phi $,
\[
(\Phi ){}^{\beta}_{\mu}(Q)=K^{A}_{\mu}(Q)
\frac{\partial {\chi}^{\beta}(Q)}{\partial Q^{A}}.
\]
  The projection operator 
\[
N^{A}_{C}\equiv{\delta}^{A}_{C}-K^{A}_{\alpha }
(\Phi ^{-1}){}^{\alpha}_{\mu}{\chi}^{\mu}_{C}
\]
 has the following properties:
\[
N^{A}_{B}N^{B}_{C}=N^{A}_{C},\,\,\,\,\,N^A_BK^B_{\mu}=0,\,\,\,\,\,
(P_{\perp})^{\tilde A}_{B}N^{C}_{\tilde A}=
(P_{\perp})^{C}_{B},\,\,\,\,\,\,\,N^{\tilde A}_
{B}(P_{\perp})^{C}_{\tilde A}=N^{C}_{B}.
\]
The matrix  ${\bar v}^{\alpha}_{\beta}(a)$ is
an inverse matrix to
 matrix ${\bar u}^{\alpha}_{\beta}(a)$.

\section{Wong's equations}
 Wong's equations can be derived from the geodesic equations written in
a special coordinate basis. So we need to change the previous coordinate basis 
$(\frac{\partial}{\partial Q^{\ast}{}^A},\frac{\partial}{\partial a^{\alpha}})$
given on $\cal P$
for a new
nonholonomic basis. This basis was introduced in our paper \cite{Storchak_3}. It
 generalizes the horizontal lift basis considered 
in \cite{Cho}. Our basis consists of the horizontal vector fields $H_A$ and the left-invariant vector fields $L_{\alpha}=v^{\mu}_{\alpha}(a)\frac{\partial}{\partial a^{\mu}}$.
The vector fields $L_{\alpha}$ obey the commutation relations
 \[
[L_{\alpha},L_{\beta}]=c^{\gamma}_{\alpha \beta} L_{\gamma},
\]
where the $c^{\gamma}_{\alpha \beta}$ are the structure constants of the group $\cal G$.

The horizontal vector fields $H_A$ are given  as follows
\[
 H_A=N^E_A(Q^{\ast}) \left(\frac{\partial}{\partial Q^{\ast}{}^E}-{\tilde {\mathscr A}}^{\alpha}_E\,L_{\alpha}\right),
\]
where ${\tilde{\mathscr A} }^{\alpha}_E(Q^{\ast},a)={\bar{\rho}}^{\alpha}_{\mu}(a)\,{\mathscr A}^{\mu}_E(Q^{\ast})$. 
The matrix ${\bar{\rho}}^{\alpha}_{\mu}$ is inverse to the matrix ${\rho}_{\alpha}^{\beta}$ of the adjoint representation of the group $\cal G$, and  ${\mathscr A}^{\nu}_P={\gamma}^{\nu\mu}K^R_{\mu}\,G_{RP}$ is the mechanical connection  defined in our principal fiber bundle.

The horizontal vector fields $H_A$ have the following commutation relation:
\[
 [H_C,H_D]=({\Lambda}^{\gamma}_CN^P_D-{\Lambda}^{\gamma}_DN^P_C)K^{S}_{{\gamma} P}\,H_S-N^E_CN^P_D\,\tilde{\mathcal F}^{\alpha}_{EP}L_{\alpha},
\]
where ${\Lambda}^{\gamma}_D=({\Phi}^{-1})^{\gamma}_{\mu}\,{\chi}^{\mu}_D$, and 
the curvature $\tilde{\mathcal F}^{\alpha}_{EP}$ of the connection ${\tilde{\mathscr A}}$ is given by
\[
\tilde{\mathcal F}^{\alpha}_{EP}=\displaystyle\frac{\partial}{\partial Q^{\ast}{}^E}\,\tilde{\mathscr A}^{\alpha}_P- 
\frac{\partial}{\partial {Q^{\ast}}^P}\,\tilde{\mathscr A}^{\alpha}_E
+c^{\alpha}_{\nu\sigma}\, \tilde{\mathscr A}^{\nu}_E\,
\tilde{\mathscr A}^{\sigma}_P,
\]
($\tilde{\mathcal F}^{\alpha}_{EP}({Q^{\ast}},a)={\bar{\rho}}^{\alpha}_{\mu}(a)\,{\mathcal F}^{\mu}_{EP}(Q^{\ast})\,$).  In the derivation of the  above commutation relation, we have used  the  equality
\[
L_{\alpha}\, {\tilde {\mathscr A}}^{\lambda}_E=-c^{\lambda}_{\alpha \mu}
\,{\tilde {\mathscr A}}^{\mu}_E.
\]
It follows from the equation satisfied by ${\rho}$:  $L_{\alpha}\,{\rho}^{\gamma}_{\beta}=c^{\mu}_{\alpha \beta}\,{\rho}^{\gamma}_{\mu}$.

In other words, the previous commutation relations represent  the commutation relations of the nonholonomic basis
\[
 [H_C,H_D]={\mathscr C}^A_{CD}\,H_A+{\mathscr C}^{\alpha}_{CD}L_{\alpha}
\]
with the structure constants 
\[{\mathscr C}^A_{CD}=({\Lambda}^{\gamma}_CK^A_{\gamma D}-{\Lambda}^{\gamma}_DK^{A}_{{\gamma} C})\]
and
\[{\mathscr C}^{\alpha}_{CD}=-N^S_CN^P_D\,\tilde{\mathcal F}^{\alpha}_{SP}\,.
\]

In our  basis, $L_{\alpha}$
commutes with $H_A$\,:
\[
[H_A,L_{\alpha}]=0.
\]

And  the  metric  (\ref{1}) has the following representation: 
\begin{equation}
\displaystyle
{\check G}_{\cal A\cal B}=
\left(
\begin{array}{cc}
G^{\rm H}_{AB} & 0 \\
0 & \tilde{\gamma }_{\alpha \beta }
\end{array}
\right),
\label{metric}
\end{equation}
where
\[
 {\tilde G}(H_A,H_B)\equiv G^{\rm H}_{AB}(Q^{\ast}), \;\;\;\; 
{\tilde{G}}(L_{\alpha},L_{\beta})\equiv\tilde{\gamma }_{\alpha \beta }(Q^{\ast},a)={\gamma}_{{\alpha}'{\beta}'}(Q^{\ast})\, {\rho}^{{\alpha}'}_{\alpha}(a)\,
{\rho}^{{\beta}'}_{\beta}(a).
\]
The ``horizontal metric'' $G^{\rm H}$ is defined by 
the projection operator ${\Pi}^{ A}_B={\delta}^A_B-K^A_{\mu}{\gamma}^{\mu \nu}K^D_{\nu}G_{DB}$ as follows:  
$G^{\rm H}_{DC}={\Pi}^{\tilde D}_D\,{\Pi}^{\tilde C}_C\,G_{{\tilde D}{\tilde C}}$.

Note that the projection operator ${\Pi}^{ A}_B$ satisfies the properties: ${\Pi}^{ A}_L N^L_C={\Pi}^{ A}_C$ and ${\Pi}^L_BN^A_L=N^A_B$.

The pseudoinverse matrix ${\check G}^{{\mathcal A}{\mathcal B}}$ to the matrix (\ref{metric}) is defined by
the following orthogonality condition:   
\begin{eqnarray*}
\displaystyle
{\check G}^{\mathcal A\mathcal B}{\check G}_{\mathcal B\mathcal C}=\left(
\begin{array}{cc}
N^A_C & 0\\
0 & {\delta}^{\alpha}_{\beta}
\end{array}
\right),
\end{eqnarray*}  
and can be written as
\begin{eqnarray*}
\displaystyle
{\check G}^{\cal A\cal B}=
\left(
\begin{array}{cc}
G^{EF}N^A_EN^B_F & 0 \\
0 & \tilde{\gamma }^{\alpha \beta }
\end{array}
\right).
\end{eqnarray*}

Using the following formula\footnote{The terms of the form  ${\partial}_{\mathcal A}\tilde G$ denote the corresponding  directional derivatives.} 
\begin{eqnarray*}
 2\,\check{\Gamma}^{{\cal D}}_{\mathcal A \mathcal B}\,{\tilde G}(
{\partial}_{\mathcal D},{\partial}_{\mathcal C})={\partial}_{\mathcal A}\,{\tilde G}({\partial}_{\mathcal B},{\partial}_{\mathcal C})+{\partial}_{\mathcal B}\,{\tilde G}({\partial}_{\mathcal A},{\partial}_{\mathcal C})-{\partial}_{\mathcal C}\,{\tilde G}({\partial}_{\mathcal A},{\partial}_{\mathcal B})& &\nonumber\\
 -{\tilde G}({\partial}_{\mathcal A},[{\partial}_{\mathcal B},{\partial}_{\mathcal C}])-{\tilde G}({\partial}_{\mathcal B},[{\partial}_{\mathcal A},{\partial}_{\mathcal C}])+{\tilde G}({\partial}_{\mathcal C},[{\partial}_{\mathcal A},{\partial}_{\mathcal B}]),
\end{eqnarray*}
 we calculated (in \cite{Storchak_3})  the Christoffel symbols
$\check{\Gamma}^{{\cal D}}_{\mathcal A \mathcal B}$ in the nonholonomic basis $(H_A,L_{\alpha})$:
\begin{eqnarray*}
&&{\check {\Gamma}}^D_{AB}=N^E_A\,{}^{\rm H}{\Gamma}^D_{BE},\nonumber\\
&&{\check {\Gamma}}^{\mu}_{AB}=-\frac12N^E_AN^F_B\,\tilde{\mathcal F}^{\mu}_{EF},\nonumber\\
&&{\check {\Gamma}}^{P}_{\alpha B}=\frac12 G^{PS}N^F_SN^E_B\,\tilde
{\mathcal F}^{\mu}_{EF}{\tilde \gamma}_{\mu\alpha},\nonumber\\
&&{\check {\Gamma}}^{P}_{A\beta}=\frac12 G^{PS}N^F_SN^E_A\,\tilde
{\mathcal F}^{\mu}_{EF}{\tilde \gamma}_{\mu\beta},\nonumber\\
&&{\check {\Gamma}}^{P}_{\alpha\beta}=-\frac12\,G^{PS}H_S{\tilde \gamma}_{\alpha \beta}=
-\frac12 G^{PS}N^E_S\,{\tilde \mathscr D}_E{\tilde \gamma}_{\alpha\beta},\nonumber\\
&&{\check{\Gamma}}^{\mu}_{\alpha B}   
=\frac12{\tilde \gamma}^{\mu\nu}H_B{\tilde \gamma}_{\alpha\nu}=
\frac12{\tilde\gamma}^{\mu\nu}N^E_B
\,{\tilde \mathscr D}_E{\tilde \gamma}_{\alpha\nu},\nonumber\\
&&{\check{\Gamma}}^{\mu}_{A\beta}   
=\frac12{\tilde \gamma}_{\mu\nu}H_A{\tilde \gamma}_{\beta\nu}=
\frac12{\tilde\gamma}^{\mu\nu}N^E_A
\,{\tilde \mathscr D}_E{\tilde \gamma}_{\beta \nu},
\nonumber\\
&&{\check{\Gamma}}^{\mu}_{\alpha\beta}=\frac12{\tilde \gamma}^{\mu\nu}(c^{\sigma}_{\alpha\beta}{\tilde \gamma}_{\sigma\nu}-c^{\sigma}_{\nu\beta}{\tilde \gamma}_{\alpha\sigma}-c^{\sigma}_{\nu\alpha}{\tilde \gamma}_{\beta\sigma}).
\end{eqnarray*}
In these formulae,
the covariant derivatives 
are given as follows
\[
{\tilde \mathscr D}_E{\tilde \gamma}_{\alpha\beta}=
\Bigl(\frac{\partial}{\partial Q^{\ast}{}^E}{\tilde \gamma}_{\alpha\beta}-c^{\sigma}_{\mu\alpha}{\tilde\mathscr A}^{\mu}_E{\tilde \gamma}_{\sigma\beta}-c^{\sigma}_{\mu\beta}{\tilde\mathscr A}^{\mu}_E{\tilde \gamma}_{\sigma\alpha}\,\Bigr).
\]

 Note also that the Christoffel symbols ${}^{\rm H}{\Gamma}^B_{CD}$ are defined by the equality
\begin{eqnarray*}
G^{\rm H}_{AB}\,
{}^{\rm H}{\Gamma}^B_{CD}
=\frac12\left(G^{\rm H}_{AC,D}+
G^{\rm H}_{AD,C}-G^{\rm H}_{CD,A}
\right).
\end{eqnarray*}
Here, 
by the derivatives we mean the following: 
$G^{\rm H}_{AC,D}\equiv 
\left.{{\partial G^{\rm H}_{AC}(Q)}\over
{\partial Q^D}}\right|_{Q=Q^{*}}$.

The geodesic equation in the Riemannian manifold $\cal P$  can be written as 
\[
 \frac{d {\dot{y}}^{\cal A}}{dt}+
{\check{\Gamma}}^{\cal A}_{\cal B \cal C}
{\dot{y}}^{\cal B}{\dot{y}}^{\cal C}=0.
\]
In our horizontal lift basis, we first decompose  the tangent vector ${\dot{y}}$  into the  horizontal and vertical components:
\[
\dot{y}(t)=z^A(t) H_A+z^{\alpha}(t)L_{\alpha}.
\]
Because of the orthogonality of our basis, we get the system of two equations:
\begin{eqnarray*}
 &&\frac{d {z}^{A}}{dt}+
{\check{\Gamma}}^{A}_{B C}\,z^Bz^C+{\check{\Gamma}}^{A}_{B \alpha}\,z^Bz^{\alpha}+{\check{\Gamma}}^{A}_{\beta B}\, z^Bz^{\beta}+{\check{\Gamma}}^{A}_{\alpha \beta}\,z^{\alpha}z^{\beta}=0
\nonumber\\
&&\frac{d {z}^{\mu}}{dt}+
{\check{\Gamma}}^{\mu}_{A B}\,z^Az^B+{\check{\Gamma}}^{\mu}_{ \alpha B}\,z^{\alpha}z^B+{\check{\Gamma}}^{\mu}_{A \beta }\, z^Az^{\beta}+{\check{\Gamma}}^{\mu}_{\alpha \beta}\,z^{\alpha}z^{\beta}=0
\nonumber\\
\end{eqnarray*}

If we replace the Christoffel symbols ${\check{\Gamma}}^{A}_{B \alpha}$ by their explicit expressions, we come to desirable equations. The horizontal equation will be as follows
\begin{equation}
 \frac{d \,{\dot {Q}^{\ast}{}^A}}{\!\!\!dt}
+{}^{\rm H}{\Gamma}^A_{BC}
{\dot {Q}^{\ast}{}^B}{\dot {Q}^{\ast}{}^C}
+G^{AS}N^F_S{\mathcal F}^{\nu}_{EF}{\dot{Q}^{\ast}{}^E}p_{\nu}
+\frac12G^{AS}\,N^E_S\,({ \mathscr D}_E{ \gamma}^{\kappa\sigma})p_{\sigma}p_{\kappa}=0.
\label{4}
\end{equation}
This equation was obtained from the previous one, after replacing  the variable $z^A$ by ${\dot {Q}^{\ast}{}^A}$. 
Also,  we have introduced a new variable $p_{\nu}={\gamma}_{\nu \kappa}{\rho}^{\kappa}_{\alpha}z^{\alpha}$.

We note that the identity $N^E_B \dot {Q}^{\ast}{}^B=\dot {Q}^{\ast}{}^A$ 
 has allowed us to omit one of the two projection operators in the third term of eq.(\ref{4}).
Moreover, in derivation of the last term of  this equation,  we have used the following identity:
\[
 ({\tilde \mathscr D}_E{\tilde \gamma}_{\beta \nu})=
-({\mathscr D}_E{\gamma}^{\kappa\sigma})\,
{\gamma}_{\mu \sigma}\,{\rho}^{\mu}_{\nu}\,
{\gamma}_{\kappa \alpha}\,{\rho}^{\alpha}_{\beta}.
\]

Making use of the method from \cite{Jadczyk}, we get the vertical Wong's equation  
\begin{equation}
 \frac{d {p}_{\sigma}}{dt}-c^{\kappa}_{\mu \sigma}\,{\mathscr A}^{\mu}_E\,p_{\kappa}\,\dot {Q}^{\ast}{}^E-c^{\kappa}_{\mu \nu}\,{\mathscr A}^{\mu}_E \,{\gamma}^{\nu \nu'}\,p_{\nu'}\,{\gamma}_{\sigma \kappa}\,\dot {Q}^{\ast}{}^E -c^{\mu}_{\sigma \nu}\,{\gamma}^{\nu \kappa}\,p_{\mu}\,p_{\kappa}=0. 
\label{5}
\end{equation}
This equation was derived by  identifying the coordinates  $z^{\alpha}$ of the vertical component  of the tangent vector ${\dot{y}}$ with the coordinates 
$u^{\alpha}_{\beta}((a(t))\frac{d a^{\beta}}{dt}$ of the vector which belongs to $T_{e}{\cal G}$.

The obtained  equation  (\ref{5}) is similar to the corresponding 
 equation from \cite{Jadczyk}. But our equation has an extra term depending on $\dot {Q}^{\ast}$, namely the third term of the equation.

\section{Principal bundle coordinates in Yang-Mills}
In Yang-Mills theory, the original evolution of the dynamical system is considered on the function space of connections that are defined  in the principal fiber  bundle, or equivalently,  on the function space of the gauge fields.   
The gauge transformations form a group which acts on this space. The reduced evolution is given on the orbit space of the group action. 

In order to apply the geometric approach, developed in \cite{Babelon}, to the study of the dynamical system defined on the space  of  connections, one must  impose additional functional restrictions both on  connections  and the gauge group  \cite{Singer_Scrip,Narasimhan,Parker,Soloviev}.
We assume that points of a manifold $\cal P$  are  the irreducible 
connections in the principal fiber bundle $P(M,G)$ (in  Sobolev class $H_{k}$, $k>3$).
Also, as the  transformation group $\cal G$, we use  the quotient group of
the gauge transformation group by its center.
Moreover, we assume that this group is the gauge group of time-independent transformations:\footnote{This means that   the gauge condition $A_0=0$ is already imposed.} 
\[
{\tilde A}^{\alpha}_i({\mathbf x})={\rho}^{\alpha}_{\beta}(g^{-1}({\mathbf x}))
{ A}^{\beta}_i({\mathbf x})+u^{\alpha}_{\mu}(g({\mathbf x}))
\frac{\partial g^{\mu}({\mathbf x})}
{\partial {\mathbf x}^i}\,,
\]
where ${\rho}^{\alpha}_{\beta}(g)=\bar{u}^{\alpha}_{\nu}(g)\,
v^{\nu}_{\beta}(g)$ is  the matrix of the adjoint
 representation of the group $G$.

To  fix the gauge symmetry, we will use  the Coulomb condition
${\partial}^k A^{\nu}_k({\mathbf x})=0$, 
$\nu=1,\dots,N_{G}$, (or ${\chi}^{\nu}(A)=0$ for short). 
This means that the original coordinates ``$Q^A$'' 
of a point $p\in \cal P$ (i.e.   the gauge fields $A^{\alpha}_i({\mathbf x})$ in our case) 
can be expressed by making use of the
coordinates of the corresponding point given on a gauge surface $\Sigma \equiv\{{\chi}^{\nu}(A)=0\}$.
As in the finite-dimensional case,  the local evolution on the orbit space corresponds to the evolution given on the surface defined by this gauge.

It is known that in the case of  time-independent gauge transformation, the Hamiltonian of the pure Yang-Mills theory, which is used in  the Schr\"odinger functional approach \cite{Rossi},  has the following form:\footnote{We restrict our consideration  to the Euclidean domain.}
\[
H=\frac 12\mu ^2\kappa \,\triangle
_{\cal P}[A_a]+\frac
1{\mu ^2\kappa }\,
V[A_a],
\]
where 
\[
\triangle _{\cal P}[A]=
\int d^3x \, k^{\alpha \beta}\delta_{ij}
\frac{{\delta}^2}{\delta A^{\alpha}_i({\mathbf x})\;
\delta A^{\beta}_j({\mathbf x})}\,,
\] 
\[
V[A]=\int d^3x \,\frac{1}{2}\,
k_{\alpha \beta}\,
F^{\alpha}_{ij}({\mathbf x})\,F^{\beta \;ij}({\mathbf x})\,.
\]
$k_{\alpha \beta}=c^{\tau}_{\mu \alpha}c^{\mu}_{\tau \beta}$ is the Cartan--Killing metric on the group G, ${\mu}^2=\hbar g_0^2$, and $\kappa$ is a real positive parameter.
This means that   for the quadratic part of the Hamiltonian,
\[
 G^{({\alpha}, i,x)\;({\beta},j,x')}
\frac{{\delta}^2}{\delta A^{(\alpha ,i,x)}\;
\delta A^{(\beta ,j,x')}},\]
one can use the flat metric 
$G^{({\alpha}, i,x)\;({\beta},j,x')}={k}^{\alpha\,\beta}\,
{\delta}^{i\,j}\,{\delta}^3({\mathbf x}-{\mathbf x}').$

Notice that in some cases, the employment of the flat metric may lead to divergences. In these cases
 one  must regularize the flat metric, i.e. convert it into a  Riemannian metric of a special form \cite{Babelon}.

In the formulae given above, we have used the extended notation for the indices from \cite{Storchak_4}.
With this notation, one can easily generalize the formulae obtained in the finite-dimensional case to the corresponding formulae  of the field theories.
 
Thus, in our problem, we have a flat Riemannian metric 
\[
ds^2=G_{(\alpha , i, x)(\beta,j,y)}\delta A^{(\alpha ,i,x)}
\delta A^{(\beta ,j,y)}\,,
\]

\[
G_{(\alpha , i, x)(\beta,j,y)}=G\biggl(\frac{\delta}
{\delta A^{\alpha}_i({\mathbf x})}\,,\,\frac{\delta}
{\delta A^{\beta}_j({\mathbf y})}\biggr)
=k_{\alpha \beta}\delta^{ij}{\delta}^3({\mathbf x}-{\mathbf y})\,,
\]
which is given on the original manifold $\cal P$ of the gauge potentials.

This manifold can be viewed locally as a total space of the principal 
fiber bundle ${\pi}: {\cal P}\to \cal M$.  It follows that  instead of the coordinates $A^{\alpha}_i({\mathbf x})$ given on the original the manifold $\cal P$ we can introduce new coordinates $(A^{\ast}{}^{\alpha}_i({\mathbf x}),g^{\mu}({\mathbf x}))$  which are related to the principal  bundle.
 The dependent coordinates $A^{\ast}$ must satisfy the equation defined by the gauge condition: $\chi ^{\alpha}(A^{\ast})=0$ .

All the transformations, that have been made ​​in the finite-dimensional case, can also be performed in the function space of the gauge fields. By using one of these transformations, i.e. the corresponding gauge transformation, we restrict  the Killing vectors $K_{(\alpha,y)}$, defined on $\cal P$ as  
\[
K_{(\alpha,y)}=K^{(\mu, i, x)}_{\;\;\;\;\;\;(\alpha,y)}\frac{\delta}
{\delta A^{(\mu ,i,x)}}\,,
\]
\[
K^{(\mu, i, x)}_{\;\;\;\;\;\;(\alpha,y)}(A)=
\left[\left({\delta}^{\;\mu}_{\alpha}{\partial}^i({\mathbf x})
+c^{\mu}_{\tilde \nu \alpha}A^{\tilde \nu i}({\mathbf x})
\right){\delta}^3 ({\mathbf x}-{\mathbf y})\right]
\equiv \left[{\mathcal D}^{\mu i}_{\;\;\alpha}(A({\mathbf x}))
\,{\delta}^3({\mathbf x}-{\mathbf y})\right]
\]
(here ${\partial}_i({\mathbf x})$ is a partial derivative 
with respect to $x^i$), to the  the gauge surface $\Sigma$.\footnote{On the surface $\Sigma$, the components $K^{(\mu, i, x)}_{\;\;\;\;\;\;(\alpha,y)}$ become dependent  on $A^{\ast}$.} 
The Killing vectors on $\Sigma$ are used for definition of the 
 orbit metric 
\[
\gamma _{(\mu,x)( \nu,y)}=K^{(\alpha,i,z)}_{\;\;\;\;\;\;(\mu,x)}
G_{(\alpha,i,z)(\beta,j,u)}K^{(\beta,j,u)}_{\;\;\;\;\;\;( \nu,y)}\,.
\]
That is, 
\[
\gamma _{(\mu,x)( \nu,y)}=\int d^3u\,d^3v\,k_{\varphi
\alpha}\,{\delta}^{kl}\,
\delta ^3({\mathbf u}-{\mathbf v})
\left[{\cal D}^{\varphi }_{\mu \,k}({\mathbf u})
\delta ^3({\mathbf u}-{\mathbf x})\right]
\left[{\cal D}^{\alpha }_{\nu \,l}({\mathbf v})
\delta ^3({\mathbf v}-{\mathbf y})\right].
\]
There is also another representation for the metric $\gamma $ : 
\[
\gamma _{(\mu,x)( \nu,y)}=k_{\varphi \alpha}{\delta}^{kl}
\left[\bigl(-\delta
^{\varphi}_{\,\mu}\;{\partial}_k({\mathbf x})+c^{\varphi}_{\sigma
\mu}A^{\ast}{}^{\sigma}_k({\mathbf x})\bigr)\bigl(\delta
^{\alpha}_{\,\nu}{\partial}_l({\mathbf y})+c^{\alpha}_{\kappa
\nu}A^{\ast}{}^{\kappa}_l({\mathbf y})\bigr){\delta}^3({\mathbf x}-{\mathbf
y})\right].
\]

An "inverse matrix" 
$\gamma ^{( \alpha,y)( \mu,z)}$ to the  matrix $\gamma _{(\mu,x)( \nu,y)}$ 
 can be defined by the following equation:
\[
\gamma _{(\mu,x)( \nu,y)}\;
\gamma ^{( \nu,y)( \sigma,z)}={\delta}^{( \sigma,z)}_
{\;(\mu,x)}\equiv{\delta}^{\sigma}_
{\,\mu}\,{\delta}^3({\mathbf z}-{\mathbf x})\,.
\]
Integrating over $\mathbf y$ in the left side of the equation (i.e. performing the generalized summation over the repeated index $y$),
we get
\[
k_{\varphi \alpha}\,{\delta}^{kl}\,
{\tilde{\cal D}}^{\varphi }_{\mu \,k}(A^{\ast}(\mathbf x))\,
{{\cal D}}^{\alpha }_{\nu \,l}(A^{\ast}(\mathbf x))
\,\gamma ^{( \nu,x)( \sigma,z)}={\delta}^{\sigma}_
{\,\mu}\,{\delta}^3(\mathbf z-\mathbf x)\,.
\]
Thus, $\gamma ^{( \nu,x)( \sigma,z)}$ is the Green function of the operator $({\tilde{\cal D}}\,{{\cal D}})_{\mu \nu}$.
Of course, this implies a  choice of the certain   boundary conditions.
We also recall that the Killing vector and the matrix ${\gamma}^{\mu\,\nu}$ are used in determining the mechanical connection 
 \[
{\mathscr A}^{\sigma}_P(Q^{\ast})={\gamma}^{\sigma\,\mu}(Q^{\ast})
\,K^R_{\mu}(Q^{\ast})\,G_{RP}(Q^{\ast})\,.
\]
Its counterpart in the Yang-Mills fields, 
the ``Coulomb connection'' 
 ${\mathscr A}^{\alpha}_{\,B}$,  is given by 
\[
{\mathscr A}^{(\alpha,x)}_{\;\;\;(\beta,j,y)}=
\left[{\cal D}^{{\varphi}}_{\;\mu j}(A^{\ast}({\mathbf y}))
{\gamma}^{(\alpha,x)\,(\mu , y)}\right]\,k_{\varphi \beta}\,.
\]

\section{Wong's equations for the gauge fields}
In this section, we  consider briefly the main steps that lead us to  Wong's equations. 
Since the  Riemannian metric on the original manifold  of the gauge fields is flat, we first rewrite
 the equation (\ref{4}),  assuming now that in the finite-dimensional case the metric is also flat, i.e. $G_{AB}={\delta}_{AB}$. 

By using the Killing relation for the flat metric, it is not difficult to find that $K^F_{\sigma}{\mathcal F}^{\nu}_{EF}=0$. This relation allows us to get rid of the projector $N$ in the equation (\ref{4}), and as a result, we come to the following equation:
\begin{equation}
 \frac{d \,{\dot {Q}^{\ast}{}^A}}{\!\!\!dt}
+{}^{\rm H}{\Gamma}^A_{BC}
{\dot {Q}^{\ast}{}^B}{\dot {Q}^{\ast}{}^C}
+G^{AS}{\mathcal F}^{\nu}_{ES}{\dot{Q}^{\ast}{}^E}p_{\nu}
+\frac12G^{AE}\,\,({ \mathscr D}_E{ \gamma}^{\kappa\sigma})p_{\sigma}p_{\kappa}=0.
\label{6}
\end{equation}

This equation will be used in derivation of the  Wong's equation for the gauge field.
 But first we must transform the terms of this equation.
We will express the  curvature ${\mathcal F}^{\nu}_{ES}$, the Christoffel symbol ${}^{\rm H}{\Gamma}^A_{BC}$ and 
${ \mathscr D}_E{ \gamma}^{\kappa\sigma}$ by using the  Killing vectors, 
 the mechanical connection and the orbit metric.

Let us consider the term of the equation with the curvature 
${\mathcal F}^{\nu}_{ES}={\mathscr A}^{\nu}_{S E}- 
{\mathscr A}^{\nu}_{E S}
+c^{\nu}_{\beta\sigma}\, {\mathscr A}^{\beta}_E\,
{\mathscr A}^{\sigma}_S.
$ 
Taking the partial derivative of the mechanical connection ${\mathscr A}^{\nu}_S={\gamma}^{\nu\mu}K^R_{\mu}\,G_{RS}$ with respect to $Q^{\ast}{}^E$, one can find, after some necessary transformations, that  
\[
{\mathcal A}^{\nu}_{SE}=-{\gamma}^{\nu \epsilon}\,K^R_{\epsilon E}\,G_{RB}\,K^B_{\mu}{\mathcal A}^{\mu}_{S}-{\mathcal A}^{\nu}_{B}\,{\mathcal A}^{\mu}_{S}\,K^B_{\mu E}+{\gamma}^{\nu \mu}\,K^B_{\mu E}\,G_{BS}.
\]
Using  the obtained  representation for ${\mathcal A}^{\nu}_{SE}$ in the curvature ${\mathcal F}^{\nu}_{ES}$, it is not difficult to get (also by means of the Killing relation for the flat metric) the following equality
\begin{eqnarray}
 &&G^{AS}{\mathcal F}^{\nu}_{ES}=2 {\gamma}^{\nu \epsilon}\,
(K^B_{\epsilon }K^R_{\mu B})\,G_{RE}\,{\gamma}^{\mu \beta}\,K^A_{\beta}+c^{\sigma}_{\mu\epsilon}\,{\gamma}^{\nu \epsilon}\,K^R_{\sigma}\,G_{RE}\,{\gamma}^{\mu \beta}\,K^A_{\beta}
\nonumber\\
&&+2{\gamma}^{\nu \mu}\,K^A_{\mu E}-2{\gamma}^{\nu \epsilon}\,(K^B_{\epsilon}K^A_{\mu B})\,{\mathcal A}^{\mu}_{E}-c^{\sigma}_{\mu\epsilon}\,{\gamma}^{\nu \epsilon}\,{\mathcal A}^{\mu}_{E}\,K^A_{\sigma}
+c^{\nu}_{\beta \sigma}\,{\gamma}^{\sigma \alpha}\,{\mathcal A}^{\beta}_{E}\,K^A_{\alpha}.
\label{7}
\end{eqnarray}
Notice that it is possible to combine the second, fifth and sixth terms  at the right hand side of eq.(\ref{7}). These terms can be rewritten to give
\[
-2{\Gamma}^{\nu}_{\kappa \sigma}{\gamma}^{\sigma \varphi}{\mathscr A}^{\kappa}_EK^A_{\varphi},
\]
in which  the Christoffel symbol  of the orbit is defined by
\[
{\Gamma}^{\nu}_{\alpha \beta}=\frac12( c^{\nu}_{\alpha\beta}-{\gamma}^{\nu \epsilon}c^{\sigma}_{\epsilon \alpha}{\gamma}_{\sigma \beta}-{\gamma}^{\nu \epsilon}c^{\sigma}_{\epsilon \beta}{\gamma}_{\sigma \alpha}).
\]

Now we consider the Christoffel symbols ${}^{\rm H}{\Gamma}^A_{BC}$ associated with  the horizontal metric ${}^{\rm H}{G}_{AB}$.
In the case of a flat  metric, it can be presented as 
\begin{equation}
 {}^{\rm H}{\Gamma}^A_{BC}=-({\mathcal A}^{\beta}_{B} 
K^A_{\beta \, C}+
{\mathcal A}^{\beta}_{C}\,K^A_{\beta  B})+\frac12(K^M_{\nu}\,K^A_{\beta M})\,({\mathcal A}^{\nu}_{B}
\,{\mathcal A}^{\beta}_{C}+{\mathcal A}^{\nu}_{C}\,{\mathcal A}^{\beta}_{B}).
\label{8}
\end{equation}
Notice  that this expression  is equal to an  analogous expression for the Christoffel symbol  obtained in \cite{Kunstatter}. 

 The  last term of eq.(\ref{6}), as can be easily shown,  has the following equivalent representation:
\begin{equation}
\frac12G^{AE}\,\,({ \mathscr D}_E{ \gamma}^{\kappa\sigma})
=(K^B_{\mu}\,K^A_{\beta B})
\,{\gamma}^{\mu \kappa}\,{\gamma}^{\beta \sigma}+c^{\kappa}_{\beta \nu}{\gamma}^{\nu \sigma}\,{\gamma}^{\beta \mu}\, K^A_{\mu}.
\label{9}
\end{equation}

In order to obtain Wong's equations for the gauge fields,
one needs to  make  a replacement of the terms of equations (\ref{7}), (\ref{8}) and (\ref{9}) by the appropriate functional expressions, and then to perform the generalized summation over the repeated  indices. 

It is easy to see, for example,  that the partial derivatives of Killing vectors in a finite-dimensional case, i.e.
\[
K^C_{\,\alpha B}(Q^{\ast})= 
\partial K^C_{\,\alpha}(Q^{\ast})/\partial Q^{\ast}{}^B\,,
\]
can be associated with the functional derivatives of $K^{(\epsilon , m,z)}_ {\;\;(\alpha,x)}$ with respect to $A^{(\beta,j,y)}$:
\[
K^{(\epsilon , m,z)}_{(\alpha,x)(\beta,j,y)}\equiv
\frac{\delta}{\delta A^{(\beta,j,y)}}\,
K^{(\epsilon , m,z)}_{\;\;\;\;(\alpha,x)} =
{\delta}^m_{\,j}\,c^{\epsilon}_{\,\beta \alpha}\,
{\delta}^3 ({\mathbf z}-{\mathbf x})\,
{\delta}^3 ({\mathbf z}-{\mathbf y})\,.
\]

Making use of the functional expression for $K^C_{\,\alpha B}$ and performing the  integration and summation over the repeated indices, we can  find    the functional representation for
the following product $K^B_{\sigma}K^E_{\alpha B}$. It is given by 
\[
K^{(\varphi, m,z)}_{(\alpha,x)(\beta,j,y)}
\,K^{(\beta,j,y)}_{(\sigma, u)}=c^{\varphi}_{\beta \alpha}
\,{\delta}^3 ({\mathbf z}-{\mathbf x})\,
\left[{\cal D}^{{\beta m}}_{\;\sigma}(A^{\ast}({\mathbf z}))
{\delta}^3 ({\mathbf z}-{\mathbf u})\right].
\] 
In other cases we will  proceed in a similar way.
Some details of these calculations are considered in  Appendix. We present here only the result of our calculation,
 according to which
the horizontal Wong's equation is as follows
\begin{eqnarray*}
&&\frac{d}{dt}\dot{A}{}^{\ast \alpha i}({\mathbf x},t)+\left(
-2\,c^{\alpha}_{\epsilon \beta}\,\dot{A}{}^{\ast \epsilon  i}({\mathbf x},t)\int d{\mathbf y}\,{\mathscr A}^{(\beta,x)}_{\;\;\;(\sigma,j,y)}\dot{A}{}^{\ast \sigma j}({\mathbf y},t)\right.\nonumber\\
&&+\left.c^{\alpha}_{\mu \beta}\int d{\mathbf y}d{\mathbf z}\,{\mathscr A}^{(\beta,x)}_{\;\;\;(\epsilon,k,z)}
\left[{\cal D}^{\mu i}_{\;\nu}(A^{\ast}({\mathbf x},t))
{\mathscr A}^{(\nu,x)}_{\;\;\;(\sigma,j,y)}
\right]\,
\dot{A}{}^{\ast \sigma j}({\mathbf y},t)\,\dot{A}{}^{\ast \epsilon k}({\mathbf z},t)\right)\nonumber\\
&&+``{\mathscr F} -{\rm terms}''\nonumber\\
&&-c^{\beta}_{\varphi \mu}\,k^{\alpha \varphi}\int d{\mathbf u}\,d{\mathbf z}\,{\gamma}^{(\kappa,u)\,(\mu, x)}
{\mathscr A}^{(\sigma, i,z)}_{\;\;\;(\beta,x)}\,p_{\kappa}({\mathbf u},t)\,p_{\sigma}({\mathbf z},t)=0,\nonumber\\
\end{eqnarray*}
where the $``{\mathscr F} -{\rm terms} ''$ corresponds to the terms of eq.(\ref{7}).  They are explicitly given as
\begin{enumerate}
\item
\[
 -2 c^{\sigma}_{\varphi \mu}k^{\beta \alpha}\int d{\mathbf y}d{\mathbf z}\,{\mathscr A}^{(\nu ,z)}_{\;\;\;(\sigma,k,y)}
\,{\mathscr A}^{(\mu,i,y)}_{\;\;\;(\beta,x)}\,\dot{A}{}^{\ast \varphi k}({\mathbf y},t)\,p_{\nu}({\mathbf z},t)
\]
\item 
\begin{eqnarray*}
&&c^{\sigma}_{\mu \epsilon}\int d{\mathbf y}\,d{\mathbf z}\,
\left\{
k_{\sigma\varphi}
\left[{\partial}_{k}({\mathbf y}){\gamma}^{(\nu,z)\,(\epsilon , y)}
\right]
k^{\beta\alpha} \,{\mathscr A}^{(\mu,i,y)}_{\;\;\;(\beta,x)}\right.
\nonumber\\
&&+\left.{\gamma}^{(\nu,z)\,(\epsilon , y)}k_{\rho\varphi}\left[ {\cal D}^{\rho}_{\;\sigma k}(A^{\ast}({\mathbf y},t))\,{\mathscr A}^{(\mu,i,y)}_{\;\;\;(\beta,x)}\right]k^{\beta\alpha}
\right\}\,\dot{A}{}^{\ast \varphi k}({\mathbf y},t)\,p_{\nu}({\mathbf z},t)
\end{eqnarray*}
\item
\[
 2\, c^{\alpha}_{\beta \mu}\left(\int d{\mathbf z}\,{\gamma}^{(\nu,z)\,(\mu , x)}\,p_{\nu}({\mathbf z},t)\right)
\dot{A}{}^{\ast \varphi i}({\mathbf x},t)
\]
\item 
\[
 2 c^{\sigma}_{\rho \mu}k^{\rho \alpha}\int d{\mathbf y}d{\mathbf z}\,{\mathscr A}^{(\nu,i,z)}_{\;\;\;(\sigma,x)}
\,{\mathscr A}^{(\mu,x)}_{\;\;\;(\varphi,k,y)}\,\dot{A}{}^{\ast \varphi k}({\mathbf y},t)\,\,p_{\nu}({\mathbf z},t)
\]
\item 
\begin{eqnarray*}
&&-c^{\sigma}_{\mu \epsilon}\int d{\mathbf y}\,d{\mathbf z}\,
\left\{
{\delta}^{\alpha}_{\sigma}
\left[{\partial}^{i}({\mathbf x}){\gamma}^{(\nu,z)\,(\epsilon , x)}
\right]
 \,{\mathscr A}^{(\mu,x)}_{\;\;\;(\varphi,k,y)}\right.
\nonumber\\
&&+\left.{\gamma}^{(\nu,z)\,(\epsilon , x)}\left[ {\cal D}^{\alpha i}_{\;\sigma }(A^{\ast}({\mathbf x},t))\,{\mathscr A}^{(\mu,x)}_{\;\;\;(\varphi,k,y)}\right]
\right\}\,\dot{A}{}^{\ast \varphi k}({\mathbf y},t)\,p_{\nu}({\mathbf z},t)
\end{eqnarray*}
\item 
\[
 c^{\nu}_{\beta \sigma}k^{\mu \alpha}\int d{\mathbf y}\,d{\mathbf z}\,{\mathscr A}^{(\beta,z)}_{\;\;\;(\varphi,k,y)}
\,{\mathscr A}^{(\sigma,i,z)}_{\;\;\;(\mu,x)}\,\dot{A}{}^{\ast \varphi k}({\mathbf y},t)\,p_{\nu}({\mathbf z},t)
\]
\end{enumerate}

The vertical Wong's equation is
\begin{eqnarray*}
&&\frac{d}{dt}\,p_{\sigma}({\mathbf x},t)-c^{\kappa}_{\varphi \sigma}\,p_{\kappa}({\mathbf x},t)\,\int d{\mathbf y}\,{\mathscr A}^{(\varphi,x)}_{\;\;\;(\beta,j,y)}\,\dot{A}{}^{\ast \beta j}({\mathbf y},t)\,\nonumber\\
&&-c^{\epsilon}_{\varphi \alpha}\,\int d{\mathbf v}\,d{\mathbf r}\,{\gamma}^{(\alpha,v)\,(\mu, r)}\,p_{\mu}({\mathbf r},t)\,{\gamma}_{(\sigma, x)\,(\epsilon , v)}\int d{\mathbf y}\,{\mathscr A}^{(\varphi,v)}_{\;\;\;(\beta,j,y)}\,\dot{A}{}^{\ast \beta j}({\mathbf y},t)\,\nonumber\\
&&-c^{\varphi}_{\sigma \epsilon}\,p_{\varphi}({\mathbf x},t)\int d{\mathbf y}\,{\gamma}^{(\epsilon , x)\,(\mu , y)}
\,p_{\mu}({\mathbf y},t)
=0.
\nonumber\\
\end{eqnarray*}

\section{Concluding remarks}
In this paper, our aim was to obtain  Wong's equations in a pure Yang--Mills theory. 
To solve this problem, we first examined a similar problem in a finite-dimensional case, using for this purpose  a special dynamical system with symmetry. The original Riemannian manifold, 
 being the configuration space of our dynamical system, can be viewed as 
a total space of the principal fiber bundle.

We have used an  approach by which the description of the evolution on the orbit space is given in terms of the dependent variables (presented in a local picture by dependent coordinates). This choice corresponds to the case of  the ``unresolved  gauges'' 
in the Yang-Mills gauge theory, and this implies that we can not find  the local coordinates  in explicit form on the gauge surfaces.

Using the bundle coordinates on the original manifold, we have derived Wong's equations 
whose solutions determine the geodesic motion in the original Riemannian manifold. 
The fact that these equations  are closely related to the reduction problems in the dynamical systems with a symmetry
allows one to use them to get analogous equations in a variety of dynamical systems with similar behavior under reduction.

In our paper,  Wong's equations in the Yang--Mills theory were obtained as a natural generalization of their finite-dimensional analogues. We found that Wong's equations are rather complicated nonlocal integro-differential equations. 

Notice that if we put the vertical variables in  Wong's equations equal to zero, we will get the equation which describes the geodesic motion on the gauge orbit space. This equation is also of interest in the gauge theories, since  it may be used there for the study of
the internal dynamics.

\section{Appendix}
Our Wong's equations for the Yang--Mills theory are obtained by  generalizing the finite-dimensional equations.
Instead of the finite-dimensional variables, we  introduce their counterparts taken from the gauge field theory.
In our new notation, the field  variables will have the extended indices. This means that now we regard the indices of the variables in the finite-dimensional equations as a compact notation of the corresponding extended indices:
\[
A\to(\alpha, i,x);\hspace{2mm}  \mu\to(\mu,u);\hspace{1mm}\ldots\hspace{1mm} {\rm etc}\;.
\]
Then for the time derivative of the basic variable $Q^B(t)$ we  have the following correspondence:
\[
{\dot {Q}^{\ast}{}^B}(t)\to \frac{d}{dt}{A}{}^{\ast (\sigma,j,y)}(t)\equiv \frac{d}{dt}{A}{}^{\ast \sigma j}({\mathbf y},t)\equiv
{\dot {A}{}^{\ast \sigma j}}({\mathbf y},t).
\]
 Thus, to obtain the Wong's equation for the gauge field we need to make  a similar replacements  in all  variables of the finite-dimensional equations. 

Let us consider the ``$\,{}^{\rm H}{\Gamma}$-terms'' of  Wong's equations 
\[
 {}^{\rm H}{\Gamma}^A_{BC}\,{\dot {Q}^{\ast}{}^B}{\dot {Q}^{\ast}{}^C}=\left(-2{\mathcal A}^{\beta}_{B} 
K^A_{\beta \, C}
+(K^M_{\nu}\,K^A_{\beta M})\,{\mathcal A}^{\nu}_{B}
\,{\mathcal A}^{\beta}_{C}\right)\,{\dot {Q}^{\ast}{}^B}{\dot {Q}^{\ast}{}^C}.
\]
We see that there are linear and quadratic  in 
${\mathcal A}$ terms on the right hand side of this equation.

The first term of $\,{}^{\rm H}{\Gamma}$ is given by $ - {\mathcal A}^{\beta}_{B} 
K^A_{\beta \, C}$. Making use of the following replacements
\[
{\mathcal A}^{\beta}_{B}\to {\mathscr A}^{(\beta,p)}_{\;\;\;(\sigma,j,y)}; \hspace{2mm}K^A_{\beta \, C}\to K^{(\alpha , i,x)}_{(\beta,p)(\epsilon,k,z)}\equiv
{\delta}^i_{\,k}\,c^{\alpha}_{\,\epsilon \beta }\,
{\delta}^3 ({\mathbf x}-{\mathbf p})\,
{\delta}^3 ({\mathbf x}-{\mathbf z})\,,
\]
we get that
\[
 {\mathcal A}^{\beta}_{B} K^A_{\beta \, C}\to \int d{\mathbf p}\,{\mathscr A}^{(\beta,p)}_{\;\;\;(\sigma,j,y)}
{\delta}^i_{\,k}\,c^{\alpha}_{\,\epsilon \beta }\,
{\delta}^3 ({\mathbf x}-{\mathbf p})\,
{\delta}^3 ({\mathbf x}-{\mathbf z})={\mathscr A}^{(\beta,x)}_{\;\;\;(\sigma,j,y)}
{\delta}^i_{\,k}\,c^{\alpha}_{\,\epsilon \beta }\,
{\delta}^3 ({\mathbf x}-{\mathbf z}).
\]
 It needs then to multiply this expression by ${\dot {A}{}^{\ast \sigma j}}({\mathbf y},t){\dot {A}{}^{\epsilon  k}}({\mathbf z},t)$ 
 and take a "generalized sum" over the repeated indices, that is
\[
 -{\delta}^i_{\,k}\int d{\mathbf y}\,d{\mathbf z}\,{\mathscr A}^{(\beta,x)}_{\;\;\;(\sigma,j,y)}
\,c^{\alpha}_{\,\epsilon \beta }\,
{\delta}^3 ({\mathbf x}-{\mathbf z}){\dot {A}{}^{\ast \sigma j}}({\mathbf y},t){\dot {A}{}^{\epsilon  k}}({\mathbf z},t).
\]
So we obtain 
\[
-c^{\alpha}_{\epsilon \beta}\,\dot{A}{}^{\ast \epsilon  i}({\mathbf x},t)\int d{\mathbf y}\,{\mathscr A}^{(\beta,x)}_{\;\;\;(\sigma,j,y)}
\dot{A}{}^{\ast \sigma j}({\mathbf y},t).
\]
This expression is the contribution to the horizontal Wong's equation arising from the first ``${\mathscr A}$-term`` of  ${}^{\rm H}{\Gamma}^A_{BC}$. 
The second term gives the same contribution. 

In Wong's equations, the ``${\mathscr A}{\mathscr A}$-terms`` of the Christoffel symbol are multiplied by a factor $(K^M_{\nu}\,K^A_{\beta M})$ which now is equal to 
\[
K^{(\mu,m,s)}_{\;\;\;(\nu, u)}\,K^{(\alpha, i,x)}_{(\beta,p)(\mu,m,s)}
=c^{\alpha}_{\mu \beta }
\,
\left[{\cal D}^{{\mu i}}_{\;\nu}(A^{\ast}({\mathbf x}))
{\delta}^3 ({\mathbf x}-{\mathbf u})\right]\, {\delta}^3 ({\mathbf x}-{\mathbf p})\,.
\] 
Because  the first ``${\mathscr A}{\mathscr A}$-term`` in $\,{}^{\rm H}{\Gamma}$ is $(K^M_{\nu}\,K^A_{\beta M}){\mathcal A}^{\nu}_{B}
\,{\mathcal A}^{\beta}_{C}$, we  first multiply the obtained  expression by ${\mathscr A}^{(\nu,u)}_{\;\;\;(\sigma,j,y)}{\mathscr A}^{(\beta,p)}_{\;\;\;(\epsilon,k,z)}$ and then integrate  it over ${\mathbf p}$ and ${\mathbf u}$. This lead us to
\begin{eqnarray*}
&&\int d{\mathbf u}\,
 c^{\alpha}_{\mu \beta }\,
\left[{\cal D}^{{\mu i}}_{\;\nu}(A^{\ast}({\mathbf x}))
{\delta}^3 ({\mathbf x}-{\mathbf u})\right]\, {\mathscr A}^{(\nu,u)}_{\;\;\;(\sigma,j,y)}{\mathscr A}^{(\beta,x)}_{\;\;\;(\epsilon,k,z)}=
\nonumber\\
&&\;\;\;\;\;\;\;\;\;\;\;\;\;\;\;\;c^{\alpha}_{\mu \beta }\,{\mathscr A}^{(\beta,x)}_{\;\;\;(\epsilon,k,z)}\left({\cal D}^{{\mu i}}_{\;\nu}(A^{\ast}({\mathbf x}))
{\mathscr A}^{(\nu,x)}_{\;\;\;(\sigma,j,y)}\right)\,.
\end{eqnarray*} 
It follows  that the contribution of the ``${\mathscr A}{\mathscr A}$-terms`` of $\,{}^{\rm H}{\Gamma}$ to Wong's equations is
\[
c^{\alpha}_{\mu \beta}\int d{\mathbf y}d{\mathbf z}\,{\mathscr A}^{(\beta,x)}_{\;\;\;(\epsilon,k,z)}
\left[{\cal D}^{\mu i}_{\;\nu}(A^{\ast}({\mathbf x},t))
{\mathscr A}^{(\nu,x)}_{\;\;\;(\sigma,j,y)}
\right]\,
\dot{A}{}^{\ast \sigma j}({\mathbf y},t)\,\dot{A}{}^{\ast \epsilon k}({\mathbf z},t)\,.
\]
Here we have taken into account the fact that the main variable ${Q}^{\ast}{}(t)$ of Wong's equations depends on the evolution parameter $t$. 
Therefore, we must replace ${A}{}^{\ast }({\mathbf y})$ by ${A}{}^{\ast }({\mathbf y},t)$ in our final formulas.

Now consider the $``{\mathscr F}$--terms'' of the horizontal  Wong's equation. In the finite-dimensional equation (\ref{6}) they are given by 
$G^{AS}{\mathcal F}^{\nu}_{ES}{\dot{Q}^{\ast}{}^E}p_{\nu}$.  In eq.(\ref{7}) we have expressed $G^{AS}{\mathcal F}^{\nu}_{ES}$ in terms of ${\mathcal A}$, $K$ and ${\gamma}^{\alpha \beta}$. The first term in this representation is 
\[
2 {\gamma}^{\nu \epsilon}\,
(K^B_{\epsilon }K^R_{\mu B})\,G_{RE}\,{\gamma}^{\mu \beta}\,K^A_{\beta}.
\]
In gauge fields, it can be rewritten as
\[
 2 {\gamma}^{(\nu,z)(\epsilon,u)}\left(\ldots \right)^{(\rho,m,v)}_{\;\;\;(\mu, v_1)(\epsilon,u)}{G}_{(\rho, m, v)(\varphi,k,y)}
{\gamma}^{(\mu,v_1)(\beta,b)}K^{(\alpha,i,x)}_{\;\;\;(\beta,b)}.
\]
Here $\left(\ldots \right)^{(\rho,m,v)}_{\;\;\;(\mu, v_1)(\epsilon,u)}$ is
\[
c^{\rho}_{\tilde \beta \mu}{\delta}^3({\mathbf v}-{\mathbf v_1})\left[{\cal D}^{\tilde \beta m}_{\;\epsilon}(A^{\ast}({\mathbf v}))
{\delta}^3({\mathbf v}-{\mathbf u})
\right],
\]
and 
 $\;\;{G}_{(\rho, m, v)(\varphi,k,y)}=k_{\rho\varphi}\,{\delta}_{mk}\,{\delta}^3({\mathbf v}-{\mathbf y})$.
Performing the integration over ${\mathbf v}$,${\mathbf v_1}$, ${\mathbf u}$ and ${\mathbf b}$, we get
\[
 2\,c^{\rho}_{\tilde \beta \mu}k_{\rho\varphi}\left[{\cal D}^{\tilde \beta }_{\;\epsilon k}(A^{\ast}({\mathbf y}))
{\gamma}^{(\nu,z)(\epsilon,y)}
\right] \Bigl[{\cal D}^{\alpha i }_{\;\beta}(A^{\ast}({\mathbf x}))
{\gamma}^{(\mu,y)(\beta,x)}
\Bigr]\,.
\]
The expression in the last bracket represents the result of the integration over ${\mathbf b}$. But 
\[
 {\mathscr A}^{(\alpha,x)}_{\;\;\;(\sigma,j,y)}k^{\sigma \nu}=\Bigl[{\cal D}^{\nu }_{\;\mu j}(A^{\ast}({\mathbf y}))
{\gamma}^{(\alpha,x)(\mu,y)}\Bigr].
\]
So, we can rewrite the obtained expression as 
\[
 2\,c^{\rho}_{\tilde \beta \mu}k_{\rho\varphi}\Bigl[{\mathscr A}^{(\nu,z)}_{\;\;\;(\sigma,k,y)}k^{\sigma \tilde \beta}\Bigr]
\Bigl[{\mathscr A}^{(\mu,i,y)}_{\;\;\;(\sigma',x)}k^{\sigma' \alpha}\Bigr].
\]
Besides, we have 
\[
c^{\rho}_{\tilde \beta \mu}k^{\sigma \tilde \beta}k_{\rho\varphi}=-c^{\sigma}_{\nu' \mu}k^{\rho \nu'}k_{\rho \varphi}=-c^{\sigma}_{\varphi \mu}.
\]
Therefore, the first term of $G^{AS}{\mathcal F}^{\nu}_{ES}$ has the following representation:
\[
 -2\,c^{\sigma}_{\varphi \mu}\,{\mathscr A}^{(\nu,z)}_{\;\;\;(\sigma,k,y)}
\,{\mathscr A}^{(\mu,i,y)}_{\;\;\;(\sigma',x)}\,k^{\sigma' \alpha}.
\]
 To obtain the first $``{\mathscr F}$--term'' of the horizontal Wong's equation one needs to multiply this expression by  
$\dot{A}{}^{\ast \varphi k}({\mathbf y},t)$ and $p_{\nu}(z,t)$ and then  perform the integration over ${\mathbf y}$ and ${\mathbf z}$.

The second term of $G^{AS}{\mathcal F}^{\nu}_{ES}$, which is given by 
\[
c^{\sigma}_{\mu\epsilon}\,{\gamma}^{\nu \epsilon}\,K^R_{\sigma}\,G_{RE}\,{\gamma}^{\mu \beta}\,K^A_{\beta},
\]
can be represented in gauge fields in the following way 
\[
c^{(\sigma, v)}_{(\mu ,v_2)(\epsilon ,u)} 
{\gamma}^{(\nu,z)(\epsilon,u)}K^{(\rho,m,v_1)}_{\;\;\;(\sigma,v)}\,{G}_{(\rho, m, v_1)(\varphi,k,y)}
{\gamma}^{(\mu,v_2)(\beta,q)}K^{(\alpha,i,x)}_{\;\;\;(\beta,q)}.
\]
Note that  $\;\;c^{(\sigma, v)}_{(\mu ,v_2)(\epsilon ,u)}  =c^{\sigma}_{\mu \epsilon }{\delta}^3({\mathbf v}-{\mathbf v_2})
{\delta}^3({\mathbf v}-{\mathbf u}).$ 

Using the explicit expressions for $K$, $G$ and $c^{(\sigma, v)}_{(\mu ,v_2)(\epsilon ,u)}$  and performing integration over ${\mathbf v}, {\mathbf u}, {\mathbf v_1},{\mathbf q}$ and  ${\mathbf v_2}$, we get
\begin{eqnarray*}
 &&c^{\sigma}_{\mu \epsilon }k_{\rho \varphi}\left\{ {\delta}^{\rho}_{\sigma} \,\biggl[{\partial}_k({\mathbf y}){\gamma}^{(\nu,z)(\epsilon,y)}\biggr]\,\biggl[
{\cal D}^{\alpha i }_{\;\beta}(A^{\ast}({\mathbf x}))
{\gamma}^{(\mu,y)(\beta,x)}\biggr]\right.
\nonumber\\
&&\;\;\;\;\;\;\;\;\;\;+\left.{\gamma}^{(\nu,z)(\epsilon,y)} {\cal D}^{\rho }_{\;\sigma k}(A^{\ast}({\mathbf y}))
\biggl[{\cal D}^{\alpha i }_{\;\beta}(A^{\ast}({\mathbf x})){\gamma}^{(\mu,y)(\beta,x)}\biggr]\right\}.
\end{eqnarray*}
Making use of the mechanical connection ${\mathscr A}$, one can  rewrite this expression  in the following form:
\begin{eqnarray*}
c^{\sigma}_{\mu \epsilon}k_{\rho\varphi}
\left\{
{\delta}^{\rho}_{\sigma}
\left[{\partial}_{k}({\mathbf y}){\gamma}^{(\nu,z)\,(\epsilon , y)}
\right]
k^{\beta\alpha} \,{\mathscr A}^{(\mu,i,y)}_{\;\;\;(\beta,x)}
+{\gamma}^{(\nu,z)\,(\epsilon , y)}\left[ {\cal D}^{\rho}_{\;\sigma k}(A^{\ast}({\mathbf y}))\,{\mathscr A}^{(\mu,i,y)}_{\;\;\;(\beta,x)}\right]k^{\beta\alpha}
\right\}\,.
\end{eqnarray*}
Notice that another way of writing this result is given by the following formula
\[
 c^{\sigma}_{\mu \epsilon}k_{\rho\varphi}\left[{\cal D}^{\rho}_{\;\sigma k}(A^{\ast}({\mathbf y}))
\left({\gamma}^{(\nu,z)\,(\epsilon , y)}{\mathscr A}^{(\mu,i,y)}_{\;\;\;(\beta,x)}\right)\right]k^{\beta\alpha}.
\]
Multiplying resulting expression by $\dot{A}{}^{\ast \varphi k}({\mathbf y},t)\,p_{\nu}({\mathbf z},t)$ and integrating over 
${\mathbf y}$ and ${\mathbf z}$, we get the second  $``{\mathscr F}$--term'' of the horizontal Wong's equation.

Finally we note that explicit expressions of other terms  of Wong's equations can be obtained by using a similar approach.

\end{document}